\documentclass{jpsj3}

\title{Suppression of Superfluidity of $^4$He in a Nanoporous Glass by Preplating a Kr Layer}

\author{Yoshiyuki Shibayama and Keiya Shirahama\thanks{E-mail address: keiya@phys.keio.ac.jp} 
}
\inst{Department of Physics, Keio University, 3-14-1 Hiyoshi, Kohoku-ku, Yokohama 223-8522 
}
\abst{Helium in nanoporous media has attracted much interest as a model Bose system with disorder and confinement. Here we have examined how a change in porous structure by preplating a monolayer of krypton affects the superfluid properties of $^4$He adsorbed or confined in a nanoporous Gelsil glass, which has a three-dimensional interconnected network of nanopores of 5.8 nm in diameter. Isotherms of adsorption and desorption of nitrogen show that monolayer preplating of Kr decreases the effective pore diameter to 4.7 nm and broadens the pore size distribution by about eight times from the sharp distribution of the bare Gelsil sample. The superfluid properties were studied by a torsional oscillator for adsorbed film states and pressurized liquid states, both before and after the monolayer Kr preplating. In the film states, both the superfluid transition temperature $T_{\mathrm c}$ and the superfluid density decrease about 10 percent by Kr preplating. The suppression of film superfluidity is attributed to the quantum localization of $^4$He atoms by the randomness in the substrate potential, which is caused by the preplating--induced broadening of the pore size distribution. In the pressurized liquid states, the superfluid density $\rho_{\mathrm s}$ is found to increase by 10 percent
 by Kr preplating, whereas $T_{\mathrm c}$ is decreased by 2 percent at all pressures. The unexpected enhancement of $\rho_{\mathrm s}$ might indicate the existence of an unknown disorder effect for confined $^4$He.}

\kword{Superfluidity, Porous Media, Localization, Helium}

\begin{document}
\maketitle

\section{Introduction}

Helium-4 ($^4$He) in porous media has attracted interest for various reasons. Porous media act as a source of disorder for an immersed $^4$He liquid and alters the universality class of the critical phenomena in the superfluid transition\cite{Chan1988}. Liquid $^4$He within porous materials undergoes a superfluid transition exhibiting critical phenomena with different critical exponents from that of bulk $^4$He, suggesting that the $^4$He--porous media system has a different universality class from the bulk. On the other hand, when a $^4$He film forms inside the pores, the dimensionality of the adsorbed $^4$He film becomes dependent on the length scale. An interesting coexistence of two- and three-dimensional features was observed in the superfluid transition\cite{Shirahama1990,BishopVycor1977}. These studies clearly demonstrate that porous media introduce a new type of disorder and spatial dimensions (2D--3D crossover and fractal dimensions) that are not realized in other Bose systems.

Recent studies using nanoporous media have revealed new phenomena in superfluid $^4$He under strong confinement\cite{Yamamoto}. When the pore size is decreased to several nanometers, the superfluidity of the confined $^4$He is strongly suppressed. The suppression becomes more prominent as the liquid pressure increases, and the superfluid transition temperature $T_{\mathrm c}$ reaches 0 K. This remarkable feature shows that confined $^4$He undergoes a quantum phase transition (QPT)\cite{QPT}. The QPT has been observed in two other regular porous materials that have different structures\cite{Taniguchi2010,Yamanaka2011}, so it is established as a ubiquitous phenomenon of confined $^4$He in nanoporous media. The emergence of QPT is related to the formation of the localized Bose--Einstein condensation (LBEC) state, in which $^4$He forms many nanometer-sized Bose condensates, and the nanoscale BECs have no global phase coherence at high temperature and high pressure\cite{Yamamoto}. The LBEC state is realized by the combined effect of the hard-core nature of $^4$He atoms and the strong confinement in nanopores. 

In general, porous media are characterized by a number of parameters (properties), such as host material, pore shape and dimensionality (tube or slit), pore connectivity, pore size and pore size distribution\cite{Porousmedia}. Among them, pore size is the most crucial parameter for the superfluid properties of confined $^4$He. A recent study has shown that the criterion for the occurrence of QPT is very sensitive to the pore size \cite{Kobayashi2010}. 

In all experiments of $^4$He in porous media performed so far, pore size was controlled by using different porous materials. It has not been possible to control the pore size during an experiment using a single porous material, and hence it has not been possible to look at subtle changes in superfluid properties when the pore size is altered. This shortcoming has prevented the development of a systematic understanding of the effect of pore size on confined $^4$He. We report here the first attempt to control the pore size of a single porous material\cite{Shibayama2007}. We have performed a torsional oscillator study of the superfluidity of $^4$He in a porous Gelsil glass, in which the effective pore size for confined liquid $^4$He is decreased by physical adsorption of a monolayer of krypton (Kr) atoms. We refer to the adsorption as 'preplating', because it is done prior to the adsorb or fill of $^4$He in the pores. The initial pore diameter of the Gelsil is 5.8 nm. Kr atoms are adsorbed on the porous glass at liquid nitrogen temperature (77 K) and form a liquid layer on the pore surface with a low vapor pressure. At lower temperatures where the superfluid study is performed, the Kr atoms form a solid layer on the pore surface with no vapor atoms. The effective pore radius for confined or adsorbed $^4$He is expected to be reduced by the thickness of the Kr layer. We have succeeded in decreasing the pore diameter to 4.7 nm by the monolayer adsorption. 

Another important characteristic of porous media is the distribution of the pore size. Although the pore size distribution can be a measure of disorder or randomness exerted on confined $^4$He, its effect on the superfluid transition has not been studied experimentally. We have found that monolayer Kr preplating broadens the pore diameter distribution by eight times from the very sharp distribution of the bare Gelsil glass. 

The changes in the two characteristics, the pore size and its distribution, result in rather unexpected behavior of the superfluid properties of confined $^4$He. In this paper we discuss the effects of Kr preplating for two different states: the adsorbed film state and the pressurized liquid state including a liquid with saturated vapor, namely the ``full pore'' case. The two states show different characteristics for the Kr preplating. The changes in the superfluid properties are discussed as effects not only of the pore size but of the broadening of its distribution. In particular, the suppression of superfluidity observed in the film states is attributed to the quantum localization of the superfluid component, i.e. Bose--Einstein condensates, by the preplating-induced random potential. A number of theoretical studies have predicted suppression of superfluidity in dilute Bose gases under a random potential\cite{Huang1993,Lopatin2002,Kobayashi2002}. The superfluid suppression can be related to disorder introduced by the Kr preplating. 

This paper is organized as follows: In Sec. \ref{Experimentals} we describe the experimental details of the Kr preplating and the torsional oscillator measurement. Section \ref{Results} discusses the experimental results. Section \ref{ResultsDistribution} details the Kr preplating study by adsorption/desorption isotherms of nitrogen (N$_2$). In Section \ref{SuperfluidProperties} we describe the basic results of the torsional oscillator measurements for the film, full pore and pressurized states. Section \ref{Discussion} is devoted to discussion, while we discuss the effect of Kr adsorption on the pore size and its distribution in \ref{KrDiscussion} and discuss the effect of Kr preplating on the superfluid transition in the two states. In Sec. \ref{Summary} we summarize the results.

\section{Experimental details}\label{Experimentals}

\subsection{Porous Gelsil Glass}\label{PGG}
We used a Gelsil glass as a porous material\cite{Gelsil}. Gelsil is a kind of gel called xerogel, which is formed by drying a SiO$_2$ gel made via a sol-gel process. It is composed mostly of SiO$_2$. Gelsil has a three-dimensional network of cylindrical nanopores. Although no systematic study has been performed on the porous structure of Gelsil, it is believed to be similar to that of Vycor, the most commonly used porous glass. The present Gelsil sample is a disk of 5.3 mm in diameter, 2.3 mm thick, and 42.85 mg in weight. The nominal pore diameter reported from the manufacturer is 5 nm. The porosity is estimated to be 0.62. As-received samples usually contain water and some hydrocarbons inside the pores, so the sample was heated at $150^{\circ} {\mathrm C}$ for one hour in a vacuum of 10$^{-6}$ Torr to remove most of the adsorbed molecules. After this degassing process, we put the Gelsil sample in a $^4$He atmosphere during the assembly of the torsional oscillator, in order to avoid completely the undesirable adsorption of water into the pores.

We obtained the surface area and pore-size distribution of the Gelsil sample from isotherm measurements for adsorption and desorption (hereafter denoted as A/D) of N$_2$ at 77 K. The results of characterization are presented in the next section.

\subsection{Preplating of Krypton}\label{Krypton}
We have chosen krypton as the preplating molecule because of its low vapor pressure at liquid nitrogen temperature, 77 K. The liquid nitrogen temperature is easiest to handle in the preplating process. The sublimation pressure $P_{\mathrm s}$ of Kr is 220 Pa at 77 K. This low pressure prevents the plugging of the gas inlet capillary (typically 0.1 mm inner diameter) in the cryostat by solidification of Kr.

We measured the A/D isotherms of both Kr and N$_2$ using a sealed glass cell containing the Gelsil sample, before installing the sample in the torsional oscillator. The cell is connected to a gas handling system that can feed both Kr and N$_2$ gases. After heating the Gelsil sample as mentioned above, the A/D isotherms of N$_2$ were measured at 77 K by immersing the cell in liquid nitrogen. The surface area of the sample was determined using the Brunauer--Emmett--Teller (BET) formula. Then, N$_2$ was pumped at room temperature and the A/D isotherms were measured for Kr. The areal density of the Kr monolayer was also determined by the BET formula. After pumping Kr atoms at room temperature, an amount of Kr corresponding to a monoatomic layer was adsorbed at 77 K. This monolayer formation took place over several days of feeding Kr gas while keeping the gas pressure below $P_{\mathrm s}$, in order to form the Kr layer as uniformly as possible. Finally, the A/D isotherms of N$_2$ were measured for the Kr preplated sample. The A/D isotherms of N$_2$ are shown in Fig. \ref{Isotherm_Fig1} and are discussed in Sec. \ref{ResultsDistribution}.

\subsection{Torsional oscillator}\label{TOstudy}
\begin{figure}[ht]
\begin{center}
\includegraphics[height=150mm]{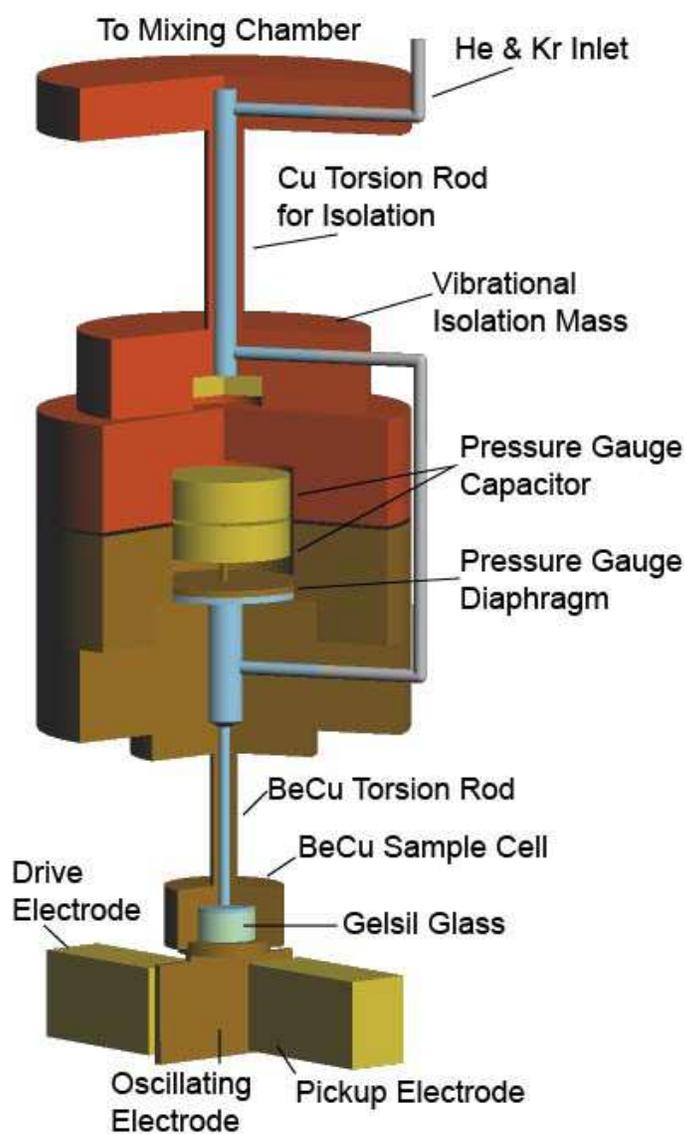}
\end{center}
\caption{Schematic view of the torsional oscillator.}
\label{TO}
\end{figure}

The torsional oscillator technique is the most powerful for studying superfluid properties of liquid helium, and it is particularly suitable for studies of $^4$He in porous media \cite{Richardson1988}. A schematic view of our oscillator is shown in Fig. \ref{TO}. The principal part is a sealed cylindrical cell containing porous material and $^4$He, which acts as the bob of a pendulum, and a hard torsion rod. The cell oscillates around the torsion axis at resonant frequency, which is determined by the torsion constant of the rod and the rotational moment of inertia of the cell. The occurrence of superfluidity in helium in the cell decreases the apparent rotational moment inertia of the cell, because the superfluid component of helium is irrotational. As a result, the resonant frequency increases below the superfluid transition temperature. 
Highly precise measurements of superfluid characteristics are realized by using a hard metal rod which has low internal loss, such as BeCu alloy. 

The sample cell is 7 mm in outer diameter and 5 mm high. The Gelsil sample is glued to the inside of a BeCu cap by Stycast 1266 epoxy resin. The torsion rod is 2 mm in outer diameter and 10 mm long. It has a 0.8 mm $\phi$ hole for the access of $^4$He into the Gelsil. The oscillator is mounted on a BeCu massive base, which acts as a vibration isolator with a second torsion rod. The base contains a diaphragm pressure gauge for measuring the pressure of the filled $^4$He. Most of these parts are machined from a single piece of BeCu. The whole setup is mounted on the mixing chamber of a dilution refrigerator. All the assemblies, from the degassing of the Gelsil sample to the mounting of the oscillator to the dilution refrigerator, are done with special care to avoid contamination by air and water. 

The oscillator motion is driven and detected by two capacitor electrodes at the bottom of the cell. A closed feedback loop circuit is used to drive the oscillator with a constant force (voltage) and to track the resonant frequency\cite{Richardson1988}. The resonant frequency $f$ (without feeding helium) is 2467.4 Hz and the quality factor $Q$ is $2\times 10^6$ at 10 mK. 

The experiment was performed down to 10 mK by a Joule--Thomson cooled dilution refrigerator. High temperatures ($1.5$--$2$ K) were realized by stopping the operation of the refrigerator.
The sample temperature $T$ was inferred by a calibrated RuO$_2$ thermometer attached to the base. 

In order to distribute $^4$He uniformly in the nanopores, a space of approximately 0.1 mm thick was opened between the Gelsil disk and the hole of the rod. Bulk liquid $^4$He contained in this open space during the experiment of pressurized states caused significant uncertainty in deriving the superfluid density in the nanopores. We will discuss this problem in Sec. \ref{PressurizedStates}. 

In torsional oscillator experiments, the superfluid transition is identified as an upward shift of the resonant frequency $f(T)$ from the so-called ``background'', which is inferred from $f(T)$ of the empty cell. The superfluid density $\rho_{\mathrm s}$, i.e. the mass fraction of the superfluid component to the total mass of $^4$He, is proportional to the shift from the background, $\Delta f(T)$.

\subsection{Procedure of measurement}\label{Procedure}
In porous media, at least two different states exhibit superfluidity: the film state and the pressurized state.
 These states are realized by controlling the amount of $^4$He fed into the pores. When the amount is small, all the $^4$He atoms are adsorbed on the pore wall and form a thin film. We refer to this state as the film state. The film state shows superfluidity at a coverage $n$, which is larger than a ``critical'' coverage $n_{\mathrm c}$. The $^4$He film formed at $n < n_{\mathrm c}$ does not show superfluidity because the $^4$He atoms are strongly bound to the pore wall. We can fill the pores with liquid $^4$He by feeding more $^4$He. This state has been customarily called the full pore state. The liquid in the full pore state has a pressure close to the saturated vapor pressure of bulk liquid $^4$He. The $^4$He density of full pore liquid can be further increased by pressurizing the liquid from room temperature. We refer to this as the pressurized state, which includes the full pore state. In this work we performed measurements at pressures up to 2.62 MPa, just below the bulk freezing pressure at 1.5 K.

The overall measurement was performed by the following procedure. The temperature dependence of the frequency $f_0(T)$ was measured before feeding any $^4$He into the cell. $f_0(T)$ was used as the temperature dependence of the empty-cell background. Then, a fixed amount of $^4$He gas was adsorbed in the cell at low temperature ($T \sim 0.1$ K). The cell was heated to 6 K for half a day in order to distribute the $^4$He uniformly in the porous glass sample. We then measured $f(T)$ and $Q(T)$ from 10 mK to about 2 K by slowly sweeping the temperature. The measurement was repeated for about 15 different $^4$He coverages. In the full pore and the pressurized states, the measurement was performed at five different pressures. After these measurements using a `bare' Gelsil, the cryostat was kept at 77 K and an amount of Kr corresponding to a monolayer was fed slowly into the cell, taking nearly a week. Then the torsional oscillator measurement was repeated for the Kr preplated sample.

\section{Experimental Results}\label{Results}
\subsection{Effect of Kr preplating on pore-size distribution}\label{ResultsDistribution}
\begin{figure}[ht]
\begin{center}
\includegraphics[height=150mm]{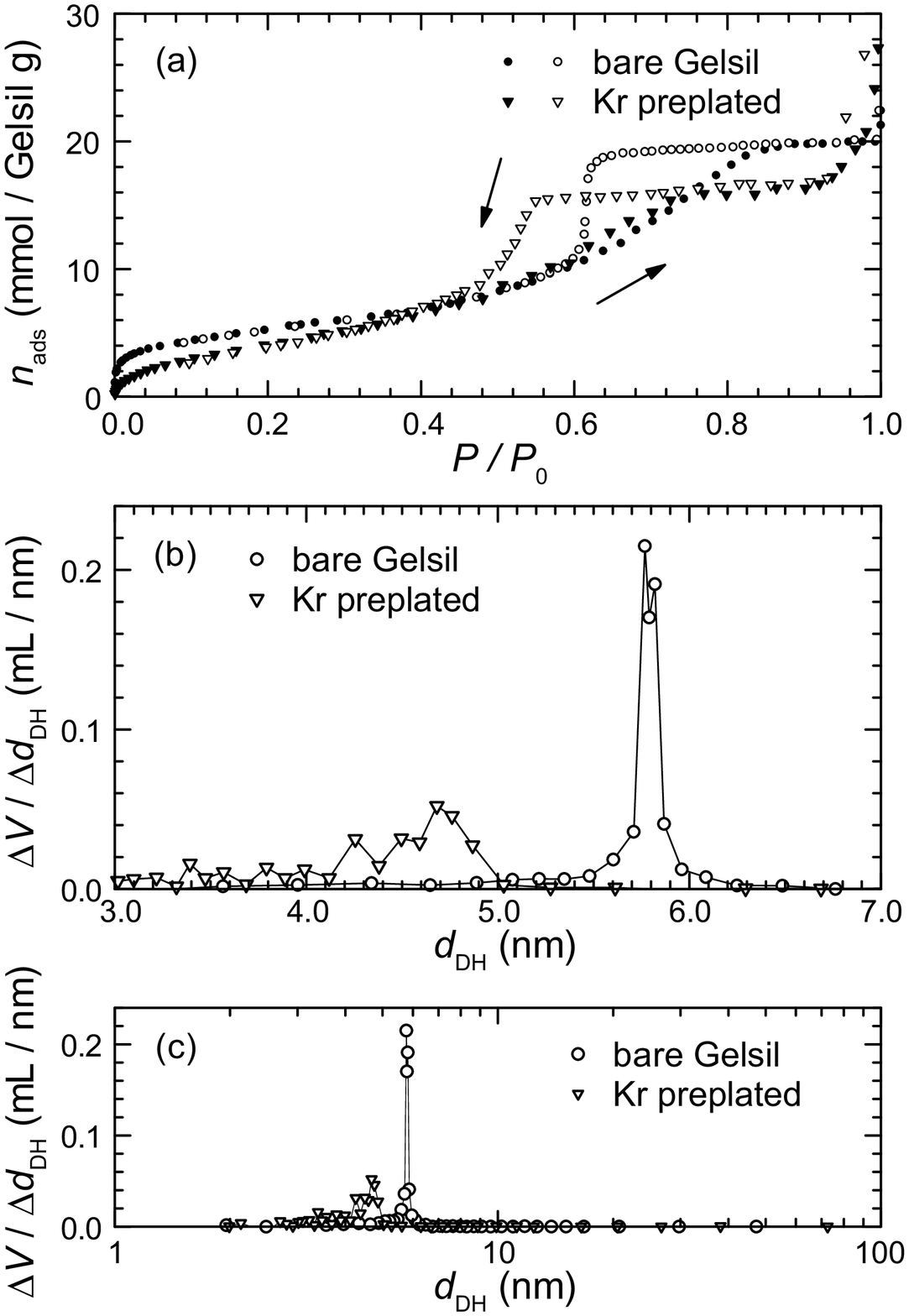}
\end{center}
\caption{(a) Isotherms of N$_2$ adsorption and desorption for the bare and Kr-preplated Gelsil. The data are molar amount of adsorbed $^4$He $n_{\mathrm {ads}}$ as a function of normalized pressure $P/P_0$, where $P_0$ denotes the saturated vapor pressure of N$_2$. The filled and open symbols are adsorption and desorption data, respectively.  (b) Pore size distribution for the bare (circle) and Kr-preplated (triangle) Gelsil obtained by the Dollimore--Heal (DH) method\cite{DH1964}. The data is plotted with a linear scale. The distribution is indicated as a derivative of pore volume $V$ with pore diameter $d_{\mathrm {DH}}$, $\Delta V / \Delta d_{\mathrm {DH}}$, as a function of $d_{\mathrm {DH}}$. (c) $\Delta V / \Delta d_{\mathrm {DH}}$ is plotted as a function of $\log d_{\mathrm {DH}}$, showing that no pores larger than 7 nm exist.}
\label{Isotherm_Fig1}
\end{figure}

Figure \ref{Isotherm_Fig1}(a) shows the A/D isotherms of N$_2$ at 77 K for the bare Gelsil sample and the sample after adsorbing Kr. The thickness of the Kr layer is estimated as $0.993 \pm 0.010$ layers from the amount of Kr admitted into the cell. 

The total pore volumes of the bare and monolayer-adsorbed Gelsil samples ($V_{\mathrm {bare}}$ and $V_{\mathrm {Kr}}$) are estimated from the isotherms to be $(32.49 \pm 0.18)\times 10^{-3}$ and $(26.0 \pm 0.2)\times 10^{-3}$ ml, respectively. The surface areas of the samples before and after Kr preplating ($S_{\mathrm {bare}}$ and $S_{\mathrm {Kr}}$) are obtained by the BET plot of the low pressure part of the data of Fig. \ref{Isotherm_Fig1}(a). $S_{\mathrm {bare}}$ and $S_{\mathrm {Kr}}$ are $(18.64 \pm 0.22) {\mathrm m}^2$ and $(17.5 \pm 0.5) {\mathrm m}^2$, from the fits for the pressure ranges $0.05 < P/P_0 < 0.30$ and $0.1 < P/P_0 < 0.30$, respectively, where $P_0$ is the saturated vapor pressure of N$_2$ at 77 K.

The isotherms shown in Fig. \ref{Isotherm_Fig1}(a) clearly indicate a difference between the bare and Kr preplated Gelsil. The hysteresis loops prove the existence of a porous structure. Before the Kr adsorption, the hysteresis loop appears in the pressure range of $0.60 < P/P_0 < 0.85$. The Kr preplating reduces the upper and lower bounds of the hysteresis to $0.35 < P/P_0 < 0.75$. Interestingly, the Kr adsorption results in broadening of the hysteresis curve and rounding of the `shoulder' in the desorption isotherm.

The reduction of the upper and lower bounds of the hysteresis indicates reduction of the effective pore size for adsorbed N$_2$. The broadening of the hysteresis curve and rounding of the desorption isotherm indicate that the distribution of the pore size is broadened. The pore diameter distribution is calculated from the desorption isotherm in Fig. \ref{Isotherm_Fig1}(a) by the Dollimore--Heal (DH) method \cite{DH1964}. 

The distributions for the bare and Kr preplated samples are shown in Fig. \ref{Isotherm_Fig1}(b) and (c).
Reflecting the steepness of the shoulder in $n_{\mathrm {ads}}$ at $P/P_0 = 0.62$, the pore diameter distribution for the bare Gelsil has a sharp single peak (maximum) at $d_{\mathrm {DH}} = 5.79$ nm. The full width at half maximum of the peak is approximately $0.019d_{\mathrm {DH}}$. 
It is therefore definitely concluded that the nanopores of the bare Gelsil are uniform in size. After the monolayer adsorption of Kr, the pore size distribution of the Kr preplated Gelsil alters to a single but much broader peak. The distribution maximum becomes located at $d_{\mathrm {DH}} = 4.8$ nm, with FWHM about $0.16d_{\mathrm {DH}}$. Thus, the monolayer Kr preplating reduces the pore diameter to about 1 nm and broadens its distribution by about eight times.

The DH method is recognized as an appropriate scheme for calculating pore diameter distribution from 3 to 30 nm. Other known methods such as the Barrett--Joyner--Halenda (BJH) method will produce similar pore size distributions with slightly different absolute values of the pore size \cite{BJH}. 
We therefore adopt the data in Fig. \ref{Isotherm_Fig1}(b) and (c) as the experimentally determined pore diameter distribution. The changes in the distribution are related to the superfluid properties of $^4$He, which will be discussed in Sec. \ref{SuperfluidProperties}.

In order to search for nanopores that are not detected by the DH analysis, we analyzed the A/D data by the so-called $\alpha _{\mathrm s}$ plot, which is sensitive to nanopores with diameter less than 2 nm \cite{alphas}. 
We found that there are no nanopores with diameter less than 2 nm in the present Gelsil sample. 

\subsection{Superfluid Properties}\label{SuperfluidProperties}
Torsional oscillator measurements reveal different superfluid characteristics for film states and pressurized states. We describe and discuss the results for these two states separately.

\subsubsection{Film States}\label{FilmStates}
\begin{figure}[ht]
\begin{center}
\includegraphics[height=160mm]{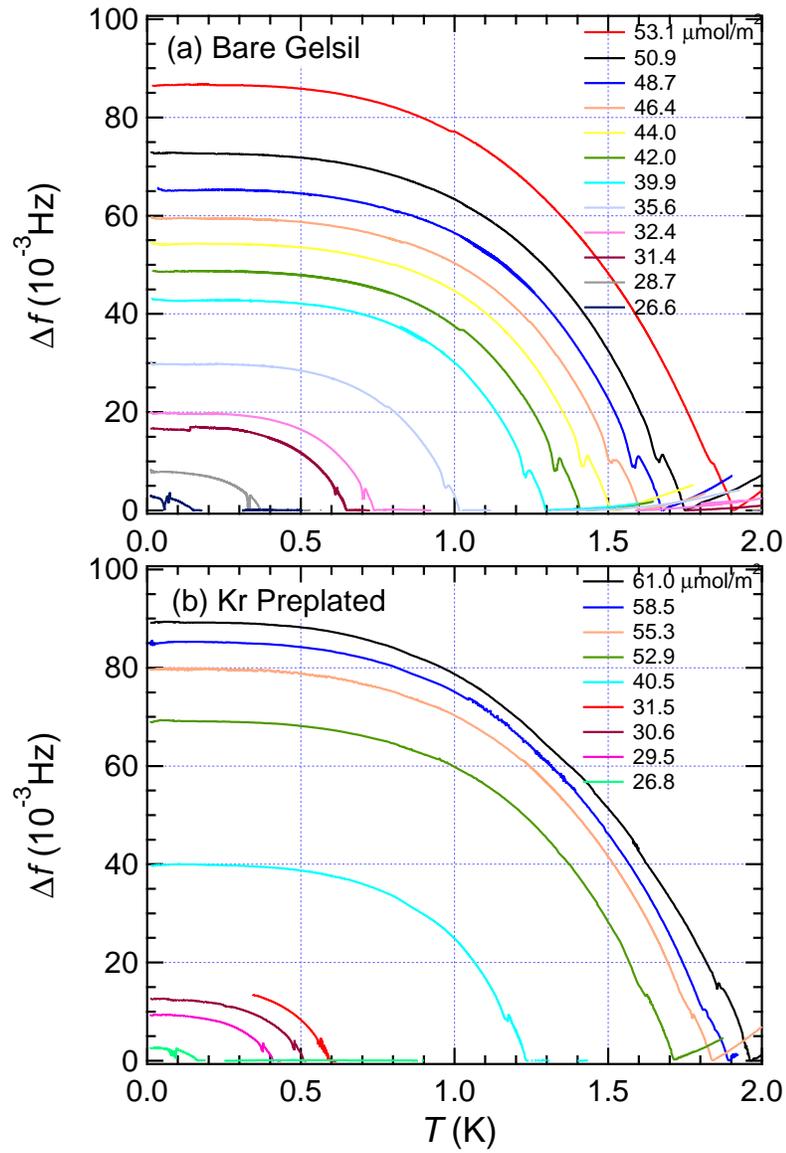}
\end{center}
\caption{Frequency shift $\Delta f(T)$ of the film states, taken at various coverages. (a) Data for bare Gelsil sample. (b) Data for Kr preplated Gelsil.}
\label{FilmDeltafColor}
\end{figure}

Figure \ref{FilmDeltafColor}(a) and (b) show a composite plot of the frequency shift $\Delta f (T)$ for about ten coverages spanning transition temperatures between 0.15 and 1.9 K. Both $\Delta f (0)$ and $T_{\mathrm c}$ increase with increasing coverage. The overall shapes of the $\Delta f (T)$ curves are approximately the same for all coverages. These features are commonly seen in $^4$He films in Vycor and other porous glasses\cite{Shirahama1990,BishopVycor1977}. A sharp dip is seen in $\Delta f (T)$ at $T \sim 0.9 T_{\mathrm c}$. This is due to the coupling of the torsional oscillation to the superfluid third sound. A rise in the frequency is seen above $T_{\mathrm c}$ in high coverage data in which $T_{\mathrm c} > 1.5$ K. This is due to the evaporation of $^4$He atoms from the adsorbed film. 

\begin{figure}[ht]
\begin{center}
\includegraphics[height=150mm]{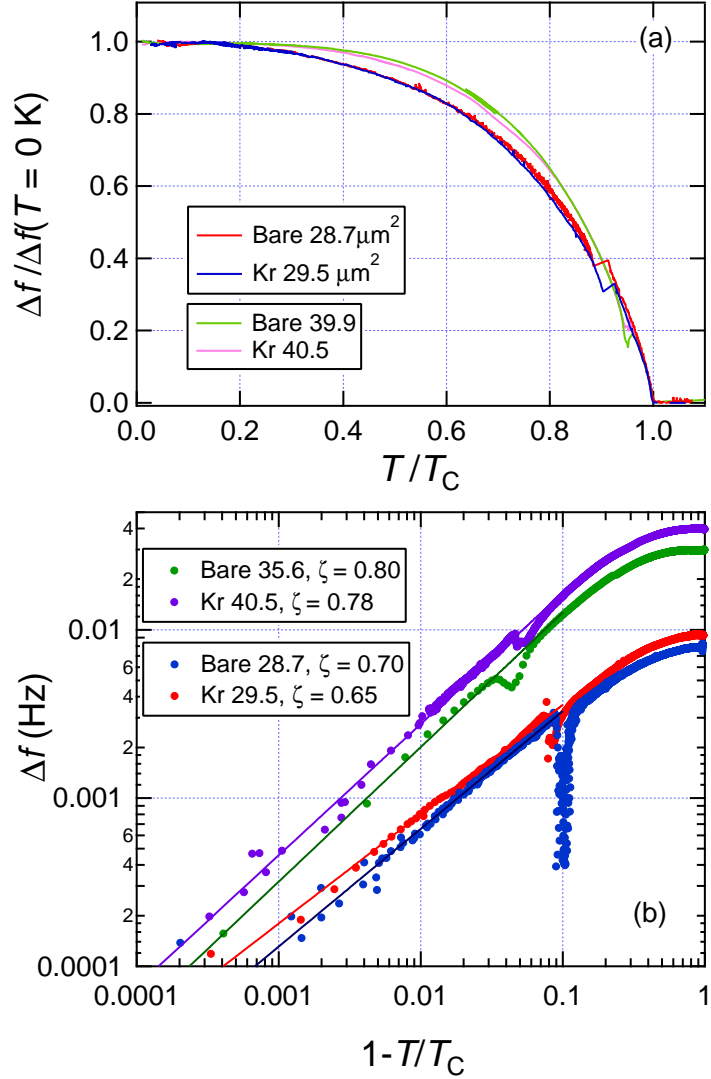}
\end{center}
\caption{(a) Frequency shift normalized by the value extrapolated to 0 K, $\Delta f(T) / \Delta f(0)$ as a function of normalized temperature $T/T_{\mathrm c}$. Two data sets with close coverages were chosen. To clarify the behavior near $T_{\mathrm c}$, spurious data points caused by the third sound resonance are eliminated. (b) $\log \Delta f(T)$ as a function of $\log (1 - T/T_{\mathrm c})$. The slope of the linear portions gives the critical exponent $\zeta$ for the superfluid density $\rho_{\mathrm s}(T) \propto (1 - T/T_{\mathrm c})^{\zeta}$.}
\label{DfNormalized}
\end{figure}

In order to examine the change in the shape of $\rho_{\mathrm s} (T)$ after Kr preplating, we plot in Fig. \ref{DfNormalized}(a) a composite of the normalized frequency shift $\Delta f(T) / \Delta f(0)$ as a function of normalized temperature $T/T_{\mathrm c}$. This quantity is proportional to the superfluid fraction $\rho_{\mathrm s} (T)/ \rho_{\mathrm s} (T=0 {\mathrm K})$. Since the shape of $\Delta f(T) / \Delta f(0)$ depends on the superfluid coverage, the data of two close coverages are plotted. It is seen that the overall shape of $\Delta f(T) / \Delta f(0)$ does not change significantly by Kr preplating. We note that the slope of the $\Delta f(T)$ curve, $d\Delta f(T)/dT$, is larger than that of $^4$He films in Vycor at low temperature ($T/T_{\mathrm c} < 0.7$, while near $T/T_{\mathrm c}$, $d\Delta f(T)/dT$ is smaller than the Vycor case\cite{BishopVycor1977}.

Two different mechanisms have been proposed for the superfluid transition of thin $^4$He films adsorbed in porous media. 
In one aspect, the three dimensional connectivity of the film and the divergence of the superfluid coherence length lead to critical behavior in the superfluid density\cite{Chan1988,BishopVycor1977,Csathy1998}. 
To examine the 3D critical property, we generate the $\log$--$\log$ plot for $\Delta f(T)$ data as a function of reduced temperature $1-T/T_{\mathrm c}$, shown in Fig. \ref{DfNormalized}(b). As in Fig. \ref{DfNormalized}(a), two pairs of data taken at close coverages are compared. The linear behavior seen in the $\log$--$\log$ plot indicates a power-law criticality with a critical exponent $\zeta$, where $\rho_{\mathrm s}(T) \propto (1 - T/T_{\mathrm c})^{\zeta}$. 
In this plot, $T_{\mathrm c}$ is determined such that the linear portion in the $\log$--$\log$ plot is as long as possible. For most of the data, the power law is observed in the temperature range $3\times 10^{-4} < 1 - T/T_{\mathrm c} < 10^{-1}$. Although the data are scattered for $1 - T/T_{\mathrm c} < 10^{-2}$, the exponent $\zeta$ is obtained by a linear fit. As is clearly seen in Fig. \ref{DfNormalized}(b), $\zeta$ depends on coverage. At low coverages, $\zeta$ is close to 0.67 (2/3), which is the value observed in bulk liquid and in thin films adsorbed on porous Vycor glass\cite{BishopVycor1977}. At high coverages where $T_{\mathrm c} > 1$ K, $\zeta$ is substantially larger than 0.75. Such a large increase in $\zeta$ at high coverages has not been reported for other porous materials (e.g. in Vycor\cite{BishopVycor1977}).

The Kr-preplated samples seem to show a slightly smaller $\zeta$ than that of the bare Gelsil data. Although this trend is seen in most of the pairs of data with close coverages, it is yet to be concluded that the difference is substantial.
To conclude, a power-law behavior is observed in the superfluid density of both bare and Kr-preplated samples, but the critical exponent and its coverage dependence are not quantitatively consistent with previous observations for $^4$He films in other porous materials\cite{Chan1988,BishopVycor1977,Csathy1998}. 

Another possible interpretation for the temperature dependence of the superfluid density is based on the theory of vortex-pair unbinding in multiply-connected $^4$He film, i.e. the Kosteritz--Thouless transition mechanism \cite{Minoguchi1989}. We will briefly discuss this mechanism in Sec. \ref{FilmDiscussion}.

\begin{figure}[ht]
\begin{center}
\includegraphics[height=170mm]{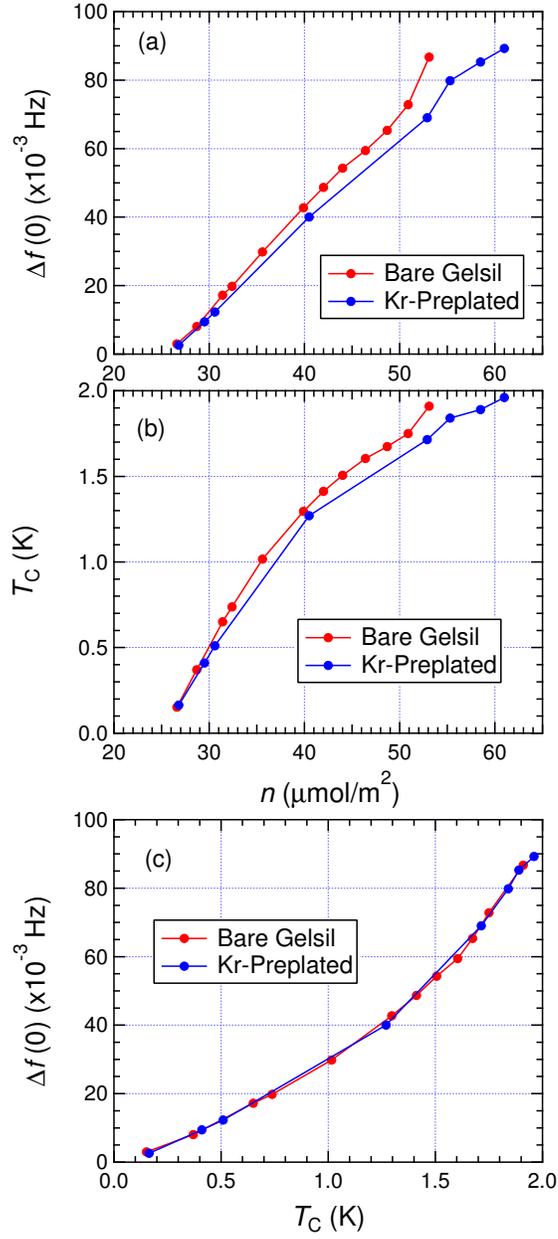}
\end{center}
\caption{Dependence of (a) $\Delta f (T = 0 {\mathrm K})$ and (b) $T_{\mathrm c}$ on coverage $n$. (c) $\Delta f (T = 0 {\mathrm K})$ versus $T_{\mathrm c}$ (combination of the data in (a) and (b)).}
\label{FilmTcDf}
\end{figure}

Next, we show the effect of Kr preplating on the coverage dependence of the superfluid density and $T_{\mathrm c}$, which is the most important finding of this work. Figures \ref{FilmTcDf} (a) and (b) show the dependencies of $\Delta f (T=0 {\mathrm K})$ ($\rho_{\mathrm s} (T=0 {\mathrm K})$) and $T_{\mathrm c}$ on coverage $n$, respectively. The critical coverage $n_{\mathrm c}$, above which the $^4$He film exhibits superfluidity, is estimated to be $25.7 \pm 0.3 \mu{\mathrm {mol/m}}^2$ for both the bare and Kr preplating samples. Within experimental accuracy, $n_{\mathrm c}$ is unchanged by the monolayer adsorption of Kr. 

In contrast, $\Delta f (0) $ and $T_{\mathrm c}$ are affected by the Kr preplating. The slopes of $\Delta f (0) $ and $T_{\mathrm c}$ against $n - n_{\mathrm c}$ decrease about 10 percent by Kr preplating. Although there is a difference in that the $\Delta f (0)$--$n$ curve is almost linear while the $T_{\mathrm c}$--$n$ curve is slightly convex, i.e. superlinear, the trend of the slopes is common. Most interestingly, the decreases in $\Delta f (0) $ and $T_{\mathrm c}$ are strongly correlated and exhibit a universal behavior. In Fig. \ref{FilmTcDf} (c) we plot $\Delta f(0)$ versus $T_{\mathrm c}$. The $\Delta f(0)$--$T_{\mathrm c}$ plots for the bare and Kr-preplated Gelsil collapse onto a single curve. Within experimental accuracy, the concave shape of the curves are identical in the range of $T_{\mathrm c}$ spanning between 0.15 and 1.90 K. All these facts clearly indicate that Kr preplating suppresses systematically the superfluidity of the adsorbed $^4$He films. We will discuss the possible mechanism for the superfluid suppression in Sec. \ref{FilmDiscussion}.

\subsubsection{Pressurized states}\label{PressurizedStates}
\begin{figure}[ht]
\begin{center}
\includegraphics[height=150mm]{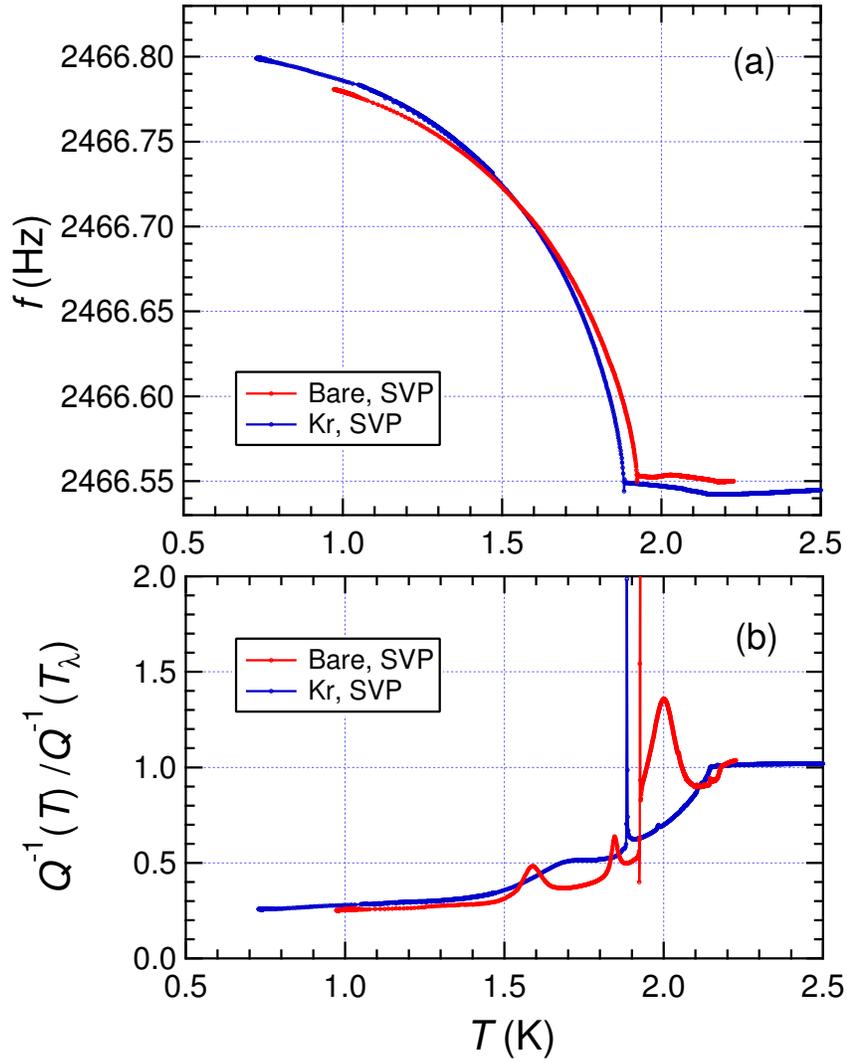}
\end{center}
\caption{Data of the full-pore state. The measurement was limited to 0.7--1.0 K due to an accidental plugging of the dilution refrigerator. (a) $f(T)$. The bulk superfluid transitions are seen at about 2.17 K, whereas the transitions in nanopores occur at lower temperatures. (b) Normalized dissipation $Q^{-1}(T)/Q^{-1}(T_{\lambda})$. A very sharp anomaly seen at $T_{\mathrm c}$ exactly coincides with the onset of the frequency shift. Other broad peaks might originate from coupling to the superfluid sound resonance modes.}
\label{FullporeSVP}
\end{figure}

In the pressurized states, the effect of Kr preplating differs from that for the film state. Figure \ref{FullporeSVP} shows the data of $f(T)$ and the dissipation normalized by the value at bulk $T_{\lambda} = 2.17 $ K, $Q^{-1}(T)/Q^{-1}(T_{\lambda})$ at SVP. The superfluid transition of $^4$He in nanopores is observed as a sharp increase in $f$ accompanied with a sharp dissipation peak. Clearly, the transition temperature $T_{\mathrm c}$ is lowered from 1.924 K to 1.884 K by Kr preplating. The sharpness of the transition is not altered by Kr preplating.

In the full-pore and pressurized measurements, the bulk liquid in the open space of the cell contributed to both $f$ and $Q^{-1}$. This prevents an exact comparison of the magnitude of the pore-origin frequency shifts for the bare Gelsil and Kr-preplated samples. However, the bulk contribution to $f(T)$ is relatively small (less than five percent of the total frequency shift) and is identical for the bare and Kr-preplated samples. Thus, we use the value $\delta f \equiv f(1.0 {\mathrm K}) - f(T_{\mathrm c})$ in place of the frequency shift originating from $^4$He in the pores.

\begin{figure}[ht]
\begin{center}
\includegraphics[height=120mm]{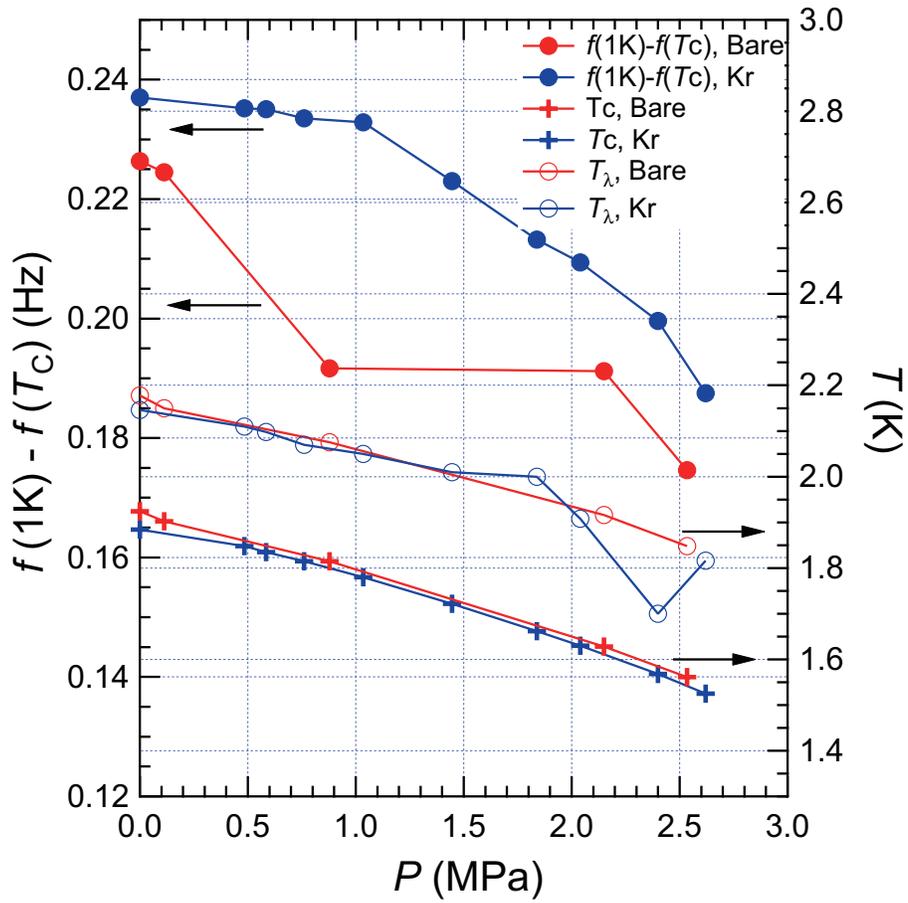}
\end{center}
\caption{Pressure dependence of $\delta f \equiv f(1.0 {\mathrm K}) - f(T_{\mathrm c})$ and $T_{\mathrm c}$. The bulk $T_{\lambda}$ obtained from $f(T)$ and $Q^{-1}(T)$ is also plotted. }
\label{Pdep}
\end{figure}

We find that $\delta f$, i.e. $\rho_{\mathrm s}$ in the pores, increases by Kr preplating. 
The enhancement of $\rho_{\mathrm s}$ is unexpected, because the reduction of $T_{\mathrm c}$ could be caused by suppression of the superfluidity. 
In Fig. \ref{Pdep} we plot $\delta f$ and $T_{\mathrm c}$.
The suppression of $T_{\mathrm c}$ and the enhancement of $\delta f$ by Kr preplating are clearly seen at all pressures.
The increase in $\delta f$ is about 10 percent and the decrease in $T_{\mathrm c}$ is 2 percent at all pressures up to 2.6 MPa.

The bare and Kr-preplated data are characterized by a very sharp superfluid onset (transition) and a concave shape of $f(T)$ just below $T_{\mathrm c}$. These features are quite similar to that for $^4$He in porous Vycor glass, in which the average pore diameter ranges from 6 to 8 nm depending on sample. This resemblance seen in the $f(T)$ curves strongly suggests that the superfluid transition of $^4$He in Gelsil can also be interpreted as a critical phenomenon obeying the power law $\rho_{\mathrm s}(T) \propto (1 - T/T_{\mathrm c})^{\zeta}$. Although it is difficult to obtain the critical exponent $\zeta$ from our raw $f(T)$ data, it is not very different from the bulk-like exponent 0.67 (2/3) that is found in $^4$He in Vycor. 

In the measurements of the pressurized state, the temperature of the sample was unfortunately limited down to 0.8 K due to plugging of the dilution refrigerator. We therefore adopt the value of $f$ at 1.0 K in the abovementioned analysis. From the behavior of $f(T)$, it is clear that the lack of data below 1 K does not influence the conclusion on the effect of Kr preplating.

\section{Discussion}\label{Discussion}
In this section, we discuss the possible effect of Kr preplating on the superfluid properties of adsorbed or confined $^4$He.
\subsection{Change in porous structure by Kr adsorption}\label{KrDiscussion}
From the pore-diameter distribution data of Fig. \ref{Isotherm_Fig1} (b) and (c), we see that the monolayer adsorption of Kr decreases the peak pore diameter from 5.8 to 4.7 nm. The van der Waals distance between two Kr Atoms and between Kr and a carbon atom in graphite are known to be 0.36 and 0.332 nm respectively. Although the distance between a Kr and a Si or O atom in Gelsil is not known, it is natural to assume that the effective diameter of a Kr atom inside the Gelsil pores is about 0.35 nm. The decrease in the pore diameter, 1.1 nm, is definitely larger than the thickness of (Kr monolayer) $\times 2$, 0.7 nm. This discrepancy is attributed to the amorphous structure of the Kr film adsorbed in nanoporous media. 
On the pore wall, the Kr adatoms do not form a spatially ordered two-dimensional solid, but form an amorphous solid that contains extra spaces between the Kr and Si or O atoms, and between neighboring Kr atoms. 
The formation of the extra volume occurs via two mechanisms: 
First, the Kr--SiO$_2$ potential energy is not spatially regular on atomic length scales. 
Second, the crystalline solid has a higher energy than the amorphous solid state, because of the lattice mismatch caused by the amorphous structure of the SiO$_2$ substrate and by the small curvature of the nanopores. 

In the amorphous Kr solid layer, the position of Kr atoms in the radial direction of the pore is not constant, but has a distribution. This can cause broadening of the pore size distribution. It is reasonable to assume a pore size distribution $\delta d \sim 0.8$ nm, which is about twice the size of a Kr atom.

\begin{figure}[ht]
\begin{center}
\includegraphics[height=100mm]{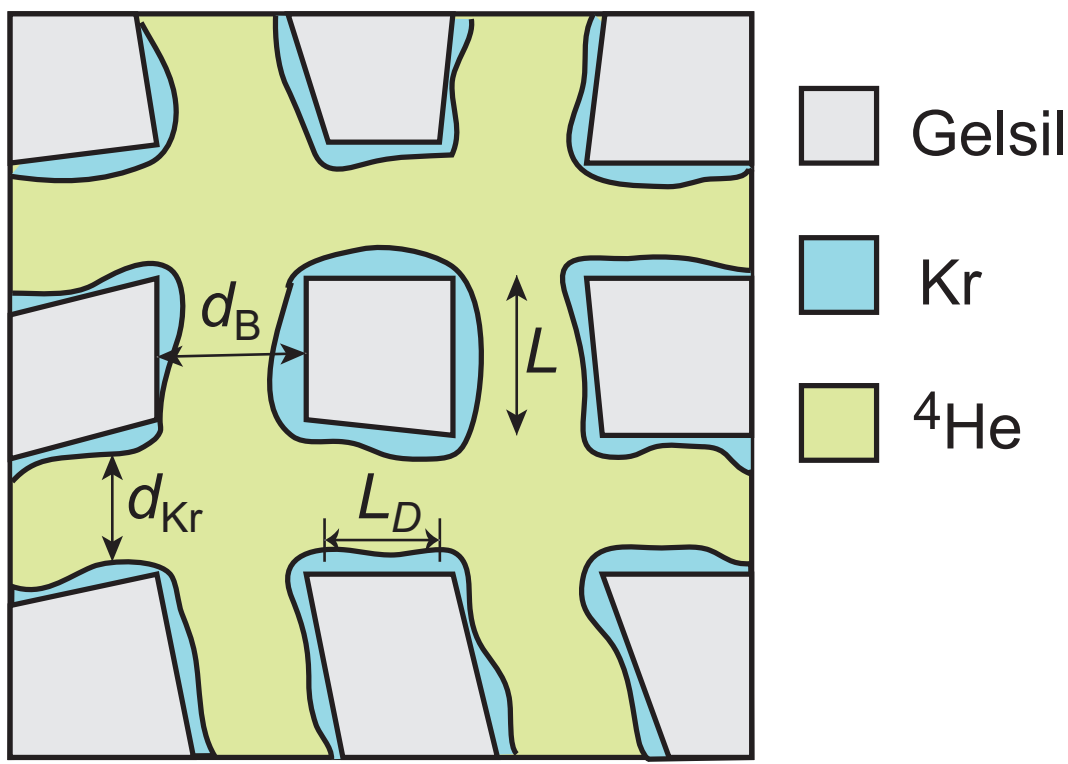}
\end{center}
\caption{Schematic of porous Gelsil and adsorbed Kr layers. The bare Gelsil sample has a quite uniform pore size $d_{\mathrm {B}}$. The average distance between the intersections of the pores, i.e. the pore length $L$, is comparable to the pore size. When a monolayer of Kr is adsorbed, the pore size $d_{\mathrm {Kr}}$ has a much broadened distribution, which produces disorder of the substrate potential. The disorder has a characteristic length scale $L_{\mathrm {D}}$, which is slightly shorter than $d_{\mathrm {B}}$ and $L$. $L_{\mathrm {D}}$ corresponds to the characteristic length in the Gaussian random potential introduced by Kobayashi and Tsubota\cite{Kobayashi2002}.}
\label{PoreSchematic}
\end{figure}

The speculated change in the pore structure is schematically illustrated in Fig. \ref{PoreSchematic}. 
The original pore size distribution of the bare Gelsil is so sharp that the pore diameter can be represented by a single value $d_{\mathrm {B}}$. 
Since the porosity is high (0.62), the average length of a pore $L$, i.e. the distance between the intersections of the pores, is comparable to $d_{\mathrm {B}}$. 
Therefore, we can speculate that the length scale of the Kr-induced additional corrugation of the substrate potential, $L_{\mathrm {D}}$, i.e. the characteristic length scale of the random potential, is comparable to or less than the pore diameter. 
This speculation can lead to a realistic functional form for the correlation function of the random potential, which has been discussed by Kobayashi and Tsubota\cite{Kobayashi2002}. 
We will later discuss this point concerning the suppression of superfluidity in the film states.

\subsection{Effect of Kr preplating on critical coverage}\label{ncdiscussion}
Preplating of monolayer Kr does not change the critical coverage $n_{\mathrm c}$. The absence of a change in $n_{\mathrm c}$ is consistent with the previous study of the preplating effect by Cs\'{a}thy et al\cite{Csathy2003}. They studied the effect of preplating four layers of various atoms on a porous gold substrate (pore diameter: 100 nm), by the torsional oscillator method. They found that $n_{\mathrm c}$ systematically decreases as the mass of the preplated atoms or molecules decrease: $n_{\mathrm c}$ for (bare Au, preplated argon, neon, and H$_2$) substrates were (25.0, 22.5, 19.0, and 6.0) $\mu {\mathrm {mol/m}}^2$, respectively. This systematic change in $n_{\mathrm c}$ was attributed to the magnitude of the depth of the $^4$He-substrate attractive potential and the magnitude of the binding energy of $^4$He atoms to the substrate.
Their result indicates that the heavier the mass of preplated atoms, the smaller the change in $n_{\mathrm c}$. 
Kr is about twice as heavy as Ar and we plated just a monolayer of Kr. It is therefore reasonable that a monolayer of Kr preplating has a negligible effect on $n_{\mathrm c}$ in our measurement.

\subsection{Superfluid suppression in film states}\label{FilmDiscussion}
The most surprising result is that both the superfluid density ($\Delta f$) and $T_{\mathrm c}$ of the films decrease by Kr preplating, while the $\Delta f(0)$--$T_{\mathrm c}$ plot collapses onto a single curve. These changes and collapse mean that Kr preplating simply decreases the number of $^4$He atoms participating in superfluidity, i.e. preplating suppresses the superfluidity of $^4$He film. 

The suppression of superfluidity is contradictory to the results of experimental and theoretical studies of the superfluid transition in $^4$He films adsorbed on porous media. Theory \cite{Minoguchi1989} shows that the superfluid response of a $^4$He film formed in multiply connected cylindrical pores is dominated by the motion of vortices (vortex--antivortex pairs) along the perimeter of a single cylindrical film.
The superfluid transition of the cylindrical $^4$He film is described by vortex--antivortex pair unbinding (the Kosterlitz--Thouless mechanism) under a periodic boundary condition due to the doubly connected nature of the cylindrical film. The transition temperature is determined by the condition that the superfluid coherence length of the film, which is the maximum distance between a vortex and an antivortex, becomes equal to half the perimeter. Since the coherence length monotonically decreases with increasing temperature, the transition temperature increases as the cylinder diameter decreases. This prediction was definitely confirmed by torsional oscillator measurements for $^4$He films adsorbed on porous glasses with various pore sizes\cite{Shirahama1990}.

However, in the present result using Gelsil, although the pore diameter is decreased from 5.8 to 4.7 nm by the monolayer Kr preplating, $T_{\mathrm c}$ decreases for the same $n - n_{\mathrm c}$. Obviously, this result cannot be explained by the abovementioned mechanism. The effect of the decrease in the pore size might be negligibly small compared to other mechanisms that suppress the superfluidity.

Here we propose a new mechanism for the superfluid suppression: The suppression is caused by the quantum localization of $^4$He atoms by a randomized substrate potential due to the Kr adatoms. Kr preplating broadens the pore diameter distribution by about eight times. This change in the pore size distribution can enhance disorder in the substrate potential. Some parts of the Bose condensate of the $^4$He film are spatially localized by the more-randomized potential. This Boson localization decreases both the superfluid density and the superfluid transition temperature. 

Bosons in random potentials and Boson localization have been theoretically studied for more than three decades. These studies are applicable to explain superfluidity of $^4$He films in porous media\cite{Hertz1979,Fisher1989}. The theories predict the suppression of superfluidity caused by the localization of Bose condensate in the random potential. 

Fisher et al. developed a scaling theory of disordered Bosons by the Bose--Hubbard model including a chemical potential that is random for each lattice site \cite{Fisher1989}. The theory lead to a number of important conclusions, e.g. the existence of a gapless Bose glass state at 0 K in the presence of randomness.
As an application of the theory, Fisher et al. discussed the behavior of $^4$He films in porous media in the vicinity of the critical coverage $n_{\mathrm {c}}$, i.e. the superfluid--``insulator'' (SI) transition. 
After this study, Crowell et al. experimentally studied the SI transition of $^4$He films in Vycor in great detail\cite{Crowell1995}. They found that Fisher's theory qualitatively explains the sublinear dependence of $\rho_{\mathrm {s}}(0)$ and $T_{\mathrm {c}}$ on $n - n_{\mathrm {c}}$. However, the insulating state, i.e. the nonsuperfluid state below $n_{\mathrm {c}}$, is not the Bose glass but a localized solid characterized by an $n$-dependent energy gap. To date, no evidence of the Bose glass has been found in thin $^4$He film. 

There is another theoretical approach that can be compared with our result. Lopatin and Vinokur\cite{Lopatin2002} obtained the phase diagram and superfluid density of a dilute Bose gas in a random potential that is characterized by a delta-function-like potential--potential correlation function, which was first introduced by Huang and Meng\cite{Huang1993}. The superfluid density and the transition temperature decreased monotonically as the degree of disorder increased\cite{Lopatin2002}. This behavior is in qualitative agreement with our observation. 

Kobayashi and Tsubota improved the potential correlation function by introducing a more realistic random potential provided by porous media, which has a Gaussian correlation with a characteristic length scale\cite{Kobayashi2002}. They found that some part of the BEC is trapped in the random potential and behaves as a normal fluid component, though Bose-condensed. The calculated superfluid density decreases with increasing randomness. The physical picture gives strong support to our proposal: The length scale of the random potential ($2\pi / k_{\mathrm p}$ in their notation) corresponds to our preplating-induced disorder length $L_{\mathrm {D}}$. The strength of the randomness ($R_0$ (Joule$^2/$m$^3$) in their notation) is also determined by the distribution of the pore size.
 
It should be emphasized that preplating-induced suppression of superfluidity was observed for the first time: The abovementioned study of the preplating effect by Cs\'{a}thy et al\cite{Csathy2003} using porous gold did not detect such a suppression. This difference is attributed to the huge difference of the pore size in the two experiments. The strength of the randomness ($R_0$ in the Kobayashi--Tsubota model\cite{Kobayashi2002}) should be scaled as $\delta d / d$, where $\delta d$ is the broadening of the pore diameter $d$ by preplating. In the Kr-preplated Gelsil $\delta d / d$ is estimated to be 0.1--0.15. The pore diameter of porous gold used by Cs\'{a}thy et al was of the order of 100 nm, which was 20 times that of our Gelsil. Even if we assumed that the four-layer preplating broadened the pore size distribution four times that of the Gelsil, $\delta d / d$ for porous gold would be estimated to be about 0.02. Thus, the effect of pore-size broadening to the randomness strength of the porous gold is negligible. This estimated ratio could be even smaller than 0.02, because it is unlikely that $\delta d$ is in proportion to the number of preplating layers. 

From the discrepancy between the Gelsil and porous gold studies, we can give a rough estimate of the length scale of disorder responsible for the superfluid suppression. The length scale is roughly equal to or less than the pore diameter $d_{\mathrm {B}}$ and the pore length $L$, which is defined as the averaged distance between the pore intersections. If disorder of atomic length scale ($\sim 0.3$ nm) was crucial for superfluid suppression, it would influence the superfluidity of $^4$He films on the preplated porous gold as well as the Gelsil case. However, no suppression of superfluidity was observed in the porous gold experiment. Thus, the characteristic length scale of disorder that is responsible for the superfluid suppression must be much larger than the atomic size. Together with the abovementioned exclusion of length scales larger than the pore size, only the randomness scale comparable to the pore size or the pore length between intersections, i.e. 4--6 nm, is relevant for the superfluid suppression. To our knowledge, this is the first possible observation of disorder-induced superfluid suppression and the first experimental estimation of the characteristic length scale of disorder relevant for the suppression. 

\subsection{Suppression of $T_{\mathrm c}$ in pressurized states}\label{TcDecreaseDiscussion}
In the full-pore and pressurized states, when $T_{\mathrm c}$ decreases about 2 percent, the frequency shift increases by the Kr preplating. 
Firstly we discuss the behavior of $T_{\mathrm c}$. At SVP, $T_{\mathrm c}$ changed from 1.924 K to 1.884 K by the change of the peak pore diameter from 5.8 to 4.7 nm.
A literature summary \cite{Brooks1979} has been published for reduced superfluid transition (onset) temperature $t_{\mathrm c} = (T_\lambda - T_{\mathrm c})/T_\lambda$ for $^4$He filled in various porous materials with pore diameters spanning between 5 and 600 nm, where $T_\lambda$ is the bulk transition temperature at SVP, 2.174 K. In our work, $T_{\mathrm c}$ is converted to $t_{\mathrm c} = 0.1150$ and 0.1334 for the bare and Kr-preplated cases, respectively. These values are located at the end of the $t_{\mathrm c}$ data set, consistent with the experimental $t_{\mathrm c}$--$d$ curve. To conclude, the suppression of $T_{\mathrm c}$ for $^4$He in bare and Kr-preplated Gelsil has the same origin as that for past experimental data of $^4$He in porous materials. As for the reduction of $T_{\mathrm c}$ in the Kr-preplated case, it is difficult to discuss the effect of a random potential, because both the decrease in the pore size and the increase in disorder can explain the slight decrease in $T_{\mathrm c}$.

A theoretical interpretation for the suppression of $T_{\mathrm c}$ in filled $^4$He in porous media has not yet been established. The abovementioned collection of $T_{\mathrm c}$ data was compared with the prediction of the so-called $\Psi$ theory\cite{Ginzburg1958,Mamaladze1967,Ginzburg1982}. The theory considers healing of the superfluid order parameter near the pore wall. $T_{\mathrm c}$ is determined by the condition that the healing length $\xi(T)$ becomes equal to the pore diameter. Mamaladze calculated $t_{\mathrm c}$ for liquid $^4$He in a cylindrical pore with diameter $d$, $t_{\mathrm c} = 2.21 \times 10^{-14} {\mathrm {m^{3/2}}} \times d^{-3/2}$\cite{Mamaladze1967,Brooks1979}. Note that the power law $t_{\mathrm c} \propto d^{-3/2}$ is the consequence of the critical behavior $\xi(T) \propto (T - T_\lambda)^{-2/3}$. However, in our opinion, although the $\Psi$ theory explains semiquantitatively the suppression of $T_{\mathrm c}$ in porous media ($5 < d < 1000$ nm), the theoretical idea severely conflicts with the existence of critical phenomenon at $T_{\mathrm c}$, where $\xi$ must diverge. Our two data sets, the bare and Kr-preplated data, are too small to judge the validity of any theoretical interpretation.

Our $T_{\mathrm c}$ data can be compared with $T_{\mathrm c}$ in other porous materials. As we described in the previous section, the overall shape of the $f(T)$ curves bears a resemblance with that for $^4$He in Vycor. We find that $T_{\mathrm c}$ at SVP in the bare Gelsil sample, 1.924 K, is slightly lower than that of full-pore $^4$He in several Vycor samples, which ranges from 1.93 to 2.05 K\cite{Kiewiet1975,Albergamo2004}. This is attributed to the difference of the pore size, in which the pore diameter of Vycor is known to span between 6 and 8 nm, depending on sample. 

Figure \ref{Pdep} shows that $T_{\mathrm c}$ decreases with increasing pressure. We see that the $T_{\mathrm c}(P)$ curve for the bare and Kr-preplated Gelsil has a slightly steeper pressure dependence than the $T_{\lambda }(P)$ curve of bulk $^4$He, particularly as the pressure is increased. Although this trend might have been seen in past torsional oscillator and ultrasound experiments for $^4$He in Vycor\cite{Beamish1983,Cao1986}, it was neither discussed nor explicitly mentioned. We believe that this change in the $T_{\mathrm c}(P)$ slope is a precursor of the quantum phase transition that has been observed in $^4$He in Gelsil with much smaller pore size ($d \sim 2.5$ nm), in which $T_{\mathrm c}$ reaches 0 K at $P \sim 3.4$ MPa\cite{Yamamoto}. A recent study using 2.5-nm Gelsil has revealed that the $T_{\mathrm c}(P)$ slope strongly depends on sample preparation (degassing of pre-adsorbed molecules such as water)\cite{Kobayashi2010}. 
A systematic study of the dependence of $T_{\mathrm c}(P)$ on various experimental parameters such as pore size and its distribution is required to understand the mechanism of superfluid suppression at high pressures.

\subsection{Enhancement of superfluid density in pressurized states}\label{PressurizedStatesDiscussion}
The increase in the superfluid density by Kr preplating is opposite from our expectation. Neither the $\Psi$ theory nor the idea of quantum localization as in the film state can explain the enhancement of $\rho_{\mathrm s}$. Besides the possibility of a novel and unknown disorder effect, we speculate that this increase is an apparent effect due to the change of the tortuosity of the porous Gelsil\cite{Bernard1991}. In torsional oscillator measurements for porous media, superfluid helium in pores whose center axes are not aligned to the tangential direction of the cylindrical cell does not perfectly decouple from the tangential oscillation of the cell. The more tortuous the pores, the more superfluid mass is coupled to the oscillation, i.e. the superfluid frequency shift decreases. The so called $\chi$ factor, which is defined as the ratio of the coupled (i.e. unobserved) superfluid mass to the total superfluid mass, introduced by Reppy and coworkers, can evaluate the tortuosity\cite{BishopKT1977}. In our film experiment, $\chi$ is estimated to be $0.70 \pm 0.03$. Although the precise evaluation of $\chi$ in the pressurized state is difficult due to the contribution of bulk $^4$He to the frequency shift, the increase in $\delta f$ suggests that $\chi$ decreases by Kr preplating. This is possible if the adsorbed Kr atoms ``smears'' the sharpness of the porous structure. The situation is schematically shown as a schematic in Fig. \ref{PoreSchematic}. The sharp edge at the crossing of the pores is smeared by Kr adatoms. This change could lead to an increase in the amount of decoupled superfluid helium in the pores in which the pore axis is perpendicular to the direction of the oscillation. As a result, $\Delta f$ may increase, and hence mask the other possible localization or size effects that decrease $\Delta f$. It is difficult to examine this idea of tortuosity change by a single experimental means (torsional oscillator) or by a single porous material. DC superflow experiments, or torsional oscillator studies using a regular nanopore array, may shed light on this problem. Also, further theoretical study for the disorder effect on superfluidity of filled $^4$He is needed.

\section{Summary and Prospects}\label{Summary}
The preplating of monolayer Kr results in two alterations to the original porous structure of the Gelsil glass. Firstly, the effective pore diameter decreases from 5.8 to 4.7 nm. 
Secondly, the pore diameter distribution increases by eight times from the sharp distribution of the bare Gelsil sample. 
The effect of Kr preplating on the superfluid properties of the adsorbed film states and the pressurized liquid states were studied by the torsional oscillator technique. In the film states, superfluidity is suppressed by the Kr preplating: the suppression is observed as a decrease in superfluid density $\rho_{\mathrm s}$ and transition temperature $T_{\mathrm c}$ by about 10 percent. The suppression of film superfluidity is attributed to the quantum localization of $^4$He atoms by the randomness in the substrate potential, which is caused by the preplating-induced broadening of the pore size distribution. In the pressurized liquid states, $T_{\mathrm c}$ decreases by 2 percent at all pressures, but $\rho_{\mathrm s}$ is found to increase by 10 percent. The decrease in $T_{\mathrm c}$ is attributed to the decrease in the pore size by the Kr preplating. The mechanism of the increase in $\rho_{\mathrm s}$ remains an open question. 

To our knowledge, the results of the film state is the first evidence of suppression of superfluidity by increasing disorder. We believe that the suppression of superfluidity by increasing randomness of the substrate potential is a general feature of $^4$He films in porous media. Although localization of Bosons has been observed in cold alkali atoms in a disordered optical potential\cite{BillyRoati2008}, macroscopic quantities such as superfluid density are difficult to extract from the visual data of cold atom systems. The preplating of Kr or other molecules can be a complementary method to dilute alkali atoms.

In order to study the confinement effect in a $^4$He-nano(meso)porous system more systematically, particularly under controlled disorder, the amorphous properties of adsorbed layers have to be better understood. This is not simply realized, but we may expect to obtain information by studying the preplating effect using porous glasses with different, especially smaller, pore sizes. The small curvature of the nanopore surface may be one of the origins of the amorphous structure of the Kr layer. Isotherm measurements of $^4$He for preplated porous media will be helpful for finding a method to control disorder.  

\section*{Acknowledgment}
This study was supported by
Grants-in-Aid for Scientific Research on Priority Area (Grant No.~17071010),
for Scientific Research (S) (Grant No.~21224010),
and for Young Scientists (B) (Grant No.~21740264)
from the Ministry of Education, Science, Sports and Culture.

\end{document}